\documentclass{PoS}
\usepackage{tabularx}
\usepackage{amsmath}
\title{Electroweak Current Operators in Chiral Effective Field Theory}

\ShortTitle{Electroweak Current Operators in chiral EFT}

\author{\speaker{Hermann Krebs}
\thanks{I would like to express my thanks to my collaborators Evgeny Epelbaum and
Ulf-G. Mei{\ss}ner for sharing their insight on the discussed
topics. I also thank the organizers for the invitation and for making this exciting
workshop possible. This work is supported by DFG (CRC110, ``Symmetries and the Emergence of Structure in QCD'').
}\\
        Ruhr-Universit\"at Bochum\\
        Institut f\"ur Theoretische Physik II\\
        E-mail: \email{hermann.krebs@rub.de}}

\abstract{In this proceeding I briefly review current status of the construction of nuclear electro-weak
currents  within chiral effective field theory. I show that gauge and chiral symmetry requirements lead
to the well-known continuity equations for the current and charge
operators which, however, get modified at higher
orders. Regularization of the current will be also
discussed. I demonstrate that implementation of a cutoff
regulator in a naive way leads to  violation  of chiral symmetry. To respect the underlying symmetries I propose to use higher derivative regularization in the nuclear forces and currents. 
}

\FullConference{The 9th International workshop on Chiral Dynamics\\
		17-21 September 2018\\
		Durham, NC, USA}

\begin{document}

\section{Nuclear Chiral Effective Field Theory}
Chiral effective field theory (EFT) is an effective field theory of quantum
chromodynamics (QCD) which works in the
energy sector where momenta of pions/nucleons are much lower than chiral
symmetry breaking scale $\Lambda_\chi\sim 1\,$GeV. Relevant symmetries
of QCD
like e.g. chiral symmetry is by construction implemented in chiral EFT in a most general
way. In the chiral EFT all the processes are described by point-like
pions and nucleons which gain their structure in a perturbative way from
loop corrections. Due to confinement  this is an efficient way to proceed since in
the low energy sector these are the observed
degrees of freedom. 

Chiral EFT has been successfully applied to meson and a nucleon
sector in entirely perturbative way. In the two- and more-nucleon
case, however, perturbative approach is not appropriate to describe
nuclei. These are bound states of nucleons which can be interpreted as
poles in the S-matrix and close to the poles any perturbation theory does
obviously not converge. Almost three decades ago, Weinberg in his
seminal papers suggested to use chiral perturbation theory to
calculate an effective interaction between nucleons (known as nuclear
forces). Bound state energies and scattering off nuclei can be
approached numerically by solving Schr\"oding equation in a
non-perturbative way~\cite{Weinberg:1991um}, see also~\cite{Epelbaum:2008ga} for a
review on this topic. This path has been followed in the last three
decade by several groups such that chiral nuclear forces have been
worked out up to next-to-next-to-next-to-next-to-leading-order
(N$^4$LO) in chiral expansion. Two-nucleon
observables calculated with N$^4$LO forces are described
with an
excellent precision~\cite{Reinert:2017usi}. 
At the same time the number of fitted parameters in
N$^4$LO forces is significantly reduced compared to phenomenological
potentials~\cite{Reinert:2017usi} which clearly underlines the importance of
two-pion-exchange contributions coming as prediction in the chiral EFT
framework. 

Within the same
formalism one can calculate nuclear electroweak current operators
consistent with the nuclear forces. The field was pioneered by Park et
al.~\cite{Park:1995pn,Park:1993jf} and was matured by two
groups who calculated leading one-loop corrections to
electroweak current operators up to N$^3$LO using two different methods to account for off-shell nuclear
effects:  the unitary transformation technique (UT) used by the
Bochum-Bonn group~\cite{Krebs:2016rqz},\cite{Krebs:2019aka} and the framework
of time-ordered perturbation theory (TOPT) used by the
Pisa-JLab group~\cite{Baroni:2015uza,Piarulli:2012bn,Pastore:2008ui,Pastore:2009is}. 

The proceeding is structured in three parts. In the first part I will briefly review our activities on the
construction of the electroweak current operator calculated within  UT
formalism. In the second part I will compare our results with the
results discussed by Pisa-Jlab group. In the third part the emphasis
will be on the symmetry preserving regulator of the current. I will
demonstrate that a naive multiplication of the current operators by a
cutoff regulator and its convolution with chiral EFT wave functions
of the deuteron leads to violation of chiral symmetry. This calls for
consistent regularization of forces and currents which preserve
underlying symmetries. Higher derivative regularization introduced by
Slavnov in the early seventies~\cite{Slavnov:1971aw} seems to be a
promising solution. 
\section{Electroweak Current Operators within UT}
Unitary transformation technique is a powerful tool to decouple
pion-nucleon and purely nucleonic states in the Fock space reducing
in this way a quantum field theoretic problem to a quantum mechanical
one. In order to formulate the problem we denote by $\lambda$ and $\eta$ projection operators which
project the states to the states with at least one pion and no pions,
respectively. The Schr\"odinger equation in the presence of external
sources can be rewritten into the
form
\begin{eqnarray}
 \left(
\begin{array}{rr}
\eta\,H\,\eta & \eta\, H\, \lambda \\
\lambda\, H\,\eta & \lambda\,H\,\lambda\\
\end{array}
\right)
\left(
\begin{array}{r}
\eta\,|\Psi\rangle\\
\lambda\,|\Psi\rangle\\
\end{array}
\right)
&=& 
i \frac{\partial}{\partial t} \left(
\begin{array}{r}
\eta\,|\Psi\rangle\\
\lambda\,|\Psi\rangle
\end{array}\right).\label{schroedinger_eq}
\end{eqnarray}
The idea is to apply a unitary transformation on the Hamilton operator $H$
in order to blockdiagonalize the matrix on the lhs of
Eq.~(\ref{schroedinger_eq}). The transformed Schr\"odinger equation
gets the form
\begin{eqnarray}
\left[U^\dagger H\,U + \left(i \frac{\partial}{\partial t}
  U^\dagger\right)U\right]U^\dagger|\Psi\rangle &=&
                                                    i\frac{\partial}{\partial t}U^\dagger|\Psi\rangle.\label{schroedinger_eq_rotated}
\end{eqnarray}
We require 
\begin{eqnarray}
&&\eta\,\left(U^\dagger H\,U\right)_s\,\lambda \,=\, \lambda\,\left(U^\dagger H\,U\right)_s\,\eta \,=\,0,\label{decoupling_eq}
\end{eqnarray}
where
\begin{eqnarray}
O_s&=&O|_{a=0, v=0, s=m_q, p=0},
\end{eqnarray}
and $O$ stays for any operator.
Here $m_q$ is a light quark mass and $a,v,s,p$ denote external axial, vector, scalar, pseudoscalar
sources, respectively. We denote a strong interacting part of the Hamiltonian by
\begin{eqnarray}
W&=&\eta \left(U^\dagger H U\right)_s \eta.
\end{eqnarray}
Note that,
although possible, we do
not require the full operator in the rectangular bracket of Eq.~(\ref{schroedinger_eq_rotated})
to be block-diagonal. It is enough that the strong interacting part of the
Hamiltonian is block diagonal (see Eq.~(\ref{decoupling_eq})). The reason is that we are not
interested in the Hamilton operator in the presence of a cloud of
external axial, vector or pseudoscalar sources. We are rather
interested in a Hamiltonian in the presence of just one (or, not in
this proceeding, maybe two) external sources. This drastically simplifies
a quantum field theoretical problem even without full
block-diagonalization.

Nuclear current operators can be extracted from first functional
derivative of the rotated Hamiltonian. In momentum space e.g. vector,
axial and pseudoscalar vector operators are defined by
\begin{eqnarray}
&&\tilde V_\mu^j(k)\,=\,\frac{\delta H_{\rm eff}}{\delta \tilde
   v_j^\mu(k)}\bigg|_s, \quad \tilde A_\mu^j(k)\,=\,\frac{\delta H_{\rm eff}}{\delta \tilde
   a_j^\mu(k)}\bigg|_s, \quad \tilde P^j(k)\,=\,\frac{\delta H_{\rm eff}}{\delta \tilde
   p_j(k)}\bigg|_s,
\end{eqnarray}
where effective Hamiltonian is 
\begin{eqnarray}
H_{\rm eff}&=&U^\dagger H\,U + \left(i \frac{\partial}{\partial t}
  U^\dagger\right)U, 
\end{eqnarray}
and Fourier transformed sources are defined by~\cite{Krebs:2016rqz}
\begin{eqnarray}
X(x)&=&\int d^4 q \,e^{-i\,q\cdot x}\tilde X(q), \quad
        X\in\left\{v_\mu^j, a_\mu^j,p^j\right\}.
\end{eqnarray}
Since $H_{\rm eff}$ is not block-diagonalized the current operators
are also not block-diagonalized which means that even if in the
initial state we have a purely nucleonic state in the final state we
can have a state with zero, one, or even more pions. However, since in
the practical calculations the currents will be convoluted with
nuclear wave functions we only need to consider purely nucleonic
initial and final states. Other states will be important e.g. if we
are interested in Compton scattering where we deal with two current
operators. In this case other matrix elements like effective pion-electroproduction
matrix-element of the vector current 
$\langle N N|V_\mu^j(k)|\pi N N\rangle$
need to be worked out. 

In the derivation of the current operator we use unitary
transformations which explicitly depend on external sources and for
this reason are time-dependent such that in general a time derivative
of the unitary transformation is non-zero. This leads to explicit
energy-transfer dependence of the currents and for this reason to a modification of
continuity equations for axial and vector current operators:
\begin{gather}
\label{continuityeqmomspacevector}
\big[W,\tilde{\bf V}_0(\vec{k},0) - \frac{\partial}{\partial k_0}\vec{k}\cdot\vec{\tilde{{\bf
  V}}}(\vec{k},k_0) + \frac{\partial}{\partial k_0}\,\big[W,\tilde{{\bf V}}_0(\vec{k},k_0)\big]\big] \,=\,\vec{k}\cdot\vec{\tilde{{\bf
  V}}}(\vec{k},0),\\
\big[W,\tilde{{\bf A}}_0(\vec{k},0) - \frac{\partial}{\partial k_0}\vec{k}\cdot\vec{\tilde{{\bf
  A}}}(\vec{k},k_0) + \frac{\partial}{\partial k_0}\,\big[W,\tilde{{\bf
   A}}_0(\vec{k},k_0)\big]+m_q \,i\frac{\partial}{\partial
   k_0}\tilde{{\bf P}}(\vec{k},k_0)\big] \nonumber\\
\,=\,\vec{k}\cdot\vec{\tilde{{\bf
  A}}}(\vec{k},0)-m_q\,i\, \tilde{{\bf P}}(\vec{k},0), 
\end{gather} 
see~\cite{Krebs:2016rqz} for derivation of these
expressions\footnote{We assume in these expressions a linear
  dependence on the energy transfer. For more complicated energy-transfer
  dependence of the currents the continuity equations look more complicated.}. Here we,
as usual,  denote by bold letters matrix elements in isospin space
\begin{eqnarray}
{\bf X}&=&\vec{X}\cdot\vec{\tau},
\end{eqnarray}
where $\tau_i$ with $i=1,2,3$ are Pauli matrices in isospin
space. Note the direct consequence form
Eq.~(\ref{continuityeqmomspacevector}) is that the knowledge of the current in the Breit frame,
where $k_0=0$ is valid, is not enough to check the continuity
equations. One needs also an information about energy-transfer
derivatives of the current operators. 

As suggested by Weinberg~\cite{Weinberg:1991um} we can use chiral
perturbation theory in order to calculate $H_{\rm eff}$. This was used
to calculate nuclear forces $W - \eta H_0 \eta$, where $H_0$ denotes a
free Hamiltonian, and nuclear current operators $V_\mu^j, A_\mu^j, P^j$. A power
counting, that tells which Feynman diagram belongs to which order in the
chiral expansion, can be derived from a naive dimensional
analysis. Denoting by
\begin{eqnarray}
Q&\sim&\{p/\Lambda_b, M_{\pi}/\Lambda_b\},
\end{eqnarray}
where $p$ stays for small momenta, $M_{\pi}$ for pion mass and
$\Lambda_b\sim 600$~MeV is a breakdown scale of the chiral expansion, we
can extract the chiral dimension $\nu$ of the corresponding diagram which
counts as $Q^\nu$ from a naive dimensional analysis.
For nuclear forces the chiral dimension of a connected diagram is given by
\begin{eqnarray}
\nu&=&-2 + \sum_i V_i \kappa_i,
\end{eqnarray}
while the chiral dimension of the nuclear charge and current operators is
given by 
\begin{eqnarray}
\label{PCkappa}
\nu=-3  + \sum_i V_i \kappa_i\,.
\end{eqnarray} 
Here $\kappa_i$ denotes inverse mass dimension of the coupling
constant at the vertex ``$i$''
and $V_i$ denotes how many times the vertex ``$i$'' appears in a considered
diagram. The inverse mass dimension can be expressed in terms of the 
chiral dimension of the vertex $d$, the number of nucleon fields $a$,
the number of pion fields $b$ and the number of external sources $c$
\begin{eqnarray}
\kappa = d + \frac{3}{2} a + b + c - 4.
\end{eqnarray}
The leading order for nuclear forces starts with one-pion-exchange and
contact interactions with $\nu=0$ and are by now calculated up to
$\nu=5$ which is N$^4$LO. Note that there are no contributions to the
nuclear forces at $\nu=1$ and next-to-leading-order (NLO)
contributions starts with $\nu=2$. For the vector and axial vector currents the leading order starts from
$\nu=-3$\footnote{For single nucleon contributions we need to subtract
three chiral dimension due to the delta function of the spectator nucleon.}. These
are the charge operator of single nucleon vector current and a current
operator of the single nucleon axial vector current. The calculations
for vector and axial vector current have been performed up to $\nu=1$
which are next-to-next-to-next-to-leading-order (N$^3$LO)
calculations. Note that similar to the nuclear forces there are no
contributions at the order $\nu=-2$ and for this reason NLO
contribution shows up first at $\nu=-1$. In
tables~\ref{tab_sum_current} and \ref{tab_sum_charge}, \ref{tab_sum_ax_current} and \ref{tab_sum_ax_charge}
all possible contributions up to N$^3$LO are summarized for vector and
axial vector operators. Note that the nucleon mass $m$ is counted as 
\begin{eqnarray}
m\sim \Lambda_b^2/p.
\end{eqnarray}
\begin{table}[t]
\caption{Chiral expansion of the nuclear electromagnetic current operator up to
  N$^3$LO. LO, NLO, N$^2$LO and N$^3$LO refer to chiral orders $Q^{-3}$,
  $Q^{-1}$, $Q^{0}$ and $Q$, respectively.  The single-nucleon contributions are given in
  Eqs.~(2.7) and (2.16) of~\cite{Krebs:2019aka}.
\label{tab_sum_current}}
\smallskip
\begin{tabularx}{\textwidth}{lcrcrcr}
\hline \hline 
  \noalign{\smallskip}
 order &&  single-nucleon  &&  two-nucleon  &&
                                                           \hspace*{1cm} three-nucleon  
\smallskip
 \\
\hline \hline 
&&&&&& \\[-7pt]
LO  && ---
             && --- && --- \\ [5pt] 
&&&&&& \\[-9pt]
NLO  && $\vec{\bf V}_{{\rm 1N:  \, static}} \;$
      
&&
                                                             $\vec{\bf
                                                             V}_{{\rm
                                                             2N:  \,
                                                             1\pi}}
                                                             $,
                                                             Eq.~(4.16)
                                          of
                                                             \cite{Kolling:2011mt} &&--- \\[9pt]
  && $+ \; \vec{\bf V}_{{\rm 1N:  \, 1/m}} \;$
   
  &&  && \\
[5pt] 
&&&&&& \\[-9pt]
N$^2$LO && $\vec{\bf V}_{{\rm 1N:  \, static}} \;$ 
&& --- && --- \\ [5pt] 
&&&&&& \\[-9pt]
N$^3$LO && $\vec{\bf V}_{{\rm 1N:  \, static}} \;$ 
&& $\vec{\bf
                                                           V}_{{\rm
                                                           2N:  \,
                                                           1\pi}} $, Eq.~(4.28)
                                          of
                                                             \cite{Kolling:2011mt}
                                            &&  ---
  \\ [2pt]
&&  $+ \; \vec{\bf V}_{{\rm 1N:  \, 1/m}}$ 
&& $+
                                                                  \; \vec{\bf
                                                           V}_{{\rm
                                                           2N:  \,
                                                           2\pi}} $,
                                                              Eq.~(2.18)
                                          of
                                                             \cite{Kolling:2009iq}  
                                            &&  
  \\ [2pt]
&& 
$+ \; \vec{\bf V}_{{\rm 1N:  \, off-shell}}$ &&
$+
                                                                  \; \vec{\bf
                                                           V}_{{\rm
                                                           2N:  \,
                                                           cont}} $, Eq.~(5.3)
                                          of
                                                             \cite{Kolling:2011mt}&&
                                                             \\ &&&&&&
                                                             \\[-7pt]
                                                             \hline \hline 
\end{tabularx}
\end{table}

\begin{table}
\caption{Chiral expansion of the nuclear electromagnetic charge operator up to
  N$^3$LO. LO, NLO, N$^2$LO and N$^3$LO refer to chiral orders $Q^{-3}$,
  $Q^{-1}$, $Q^{0}$ and $Q$, respectively. The single-nucleon contributions are given in
  Eq.~(2.6) of~\cite{Krebs:2019aka}.
\label{tab_sum_charge}}
\smallskip
\begin{tabularx}{\textwidth}{lcrrr}
  \hline \hline 
\noalign{\smallskip}
 order &&  single-nucleon  &  two-nucleon  &
                                                         three-nucleon  
\smallskip
 \\
\hline \hline 
&&&& \\[-7pt]
LO && ${\bf V}^0_{{\rm 1N:  \, static}} \;$ 
             & --- 
& --- 
\\ [5pt] 
&&&& \\[-9pt]
NLO &&  ${\bf V}^0_{{\rm 1N:  \, static}} \;$
                           &--- 
& --- \\
[5pt] 
&&&& \\[-9pt]
N$^2$LO && ${\bf V}^0_{{\rm 1N:  \, static}} \;$ 
                   & --- 
& --- 
\\ [5pt] 
&&&& \\[-9pt]
N$^3$LO && ${\bf V}^0_{{\rm 1N:  \, static}} \;$ 
            & ${\bf
                                                           V}^0_{{\rm
                                                           2N:  \,
                                                           1\pi}} $, Eq.~(4.30)
                                          of
                                                             \cite{Kolling:2011mt}
                                            &  ${\bf V}^0_{{\rm 3N:
                                              \, \pi}}$,
                                              Eq.~(4.1)
                                              of~\cite{Krebs:2019aka} 
  \\ [2pt]
&&  
$+ \; {\bf V}^0_{{\rm 1N:  \, 1/m}}$ 
& 
$+
                                                                  \; {\bf
                                                           V}^0_{{\rm
                                                           2N:  \,
                                                           2\pi}} $,
                                                              Eq.~(2.19)
                                          of
                                                             \cite{Kolling:2009iq}   
                                            &   $+ \; {\bf V}^0_{{\rm 3N:
                                              \, \pi}}$,
                                              Eq.~(4.2)
                                              of~\cite{Krebs:2019aka} 
  \\ [2pt]
&&$+ \; {\bf V}^0_{{\rm 1N:  \, 1/m^2}}$ 

 &
 $+
                                                                  \; {\bf
                                                           V}^0_{{\rm
                                                           2N:  \,
                                                           cont}} $, Eq.~(5.6)
                                          of
                                                             \cite{Kolling:2011mt}
& $+ \; {\bf V}^0_{{\rm 3N:
                                              \, cont}}$,
                                              Eq.~(4.3)  of~\cite{Krebs:2019aka}
  \\
[2pt]
&&  &
 $+
                                                                  \; {\bf
                                                           V}^0_{{\rm
                                                           2N:  \,
                                                           1\pi , \, 1/m}} $, Eq.~(4.30)
                                          of
                                                             \cite{Kolling:2011mt}&
                                                             \\
                                                             &&&& \\[-7pt]
                                                             \hline \hline 
\end{tabularx}
\end{table}

\begin{table}
\caption{Chiral expansion of the nuclear axial current operator up to
  N$^3$LO. LO, NLO, N$^2$LO and N$^3$LO refer to chiral orders $Q^{-3}$,
  $Q^{-1}$, $Q^{0}$ and $Q$, respectively.  All equation references are understood to be from~\cite{Krebs:2016rqz}. 
\label{tab_sum_ax_current}}
\smallskip
\begin{tabularx}{\textwidth}{lrrr}
\noalign{\smallskip}
\hline \hline
 order &  single-nucleon  &  two-nucleon  &
                                                            three-nucleon  
\smallskip
 \\
\hline \hline
&&& \\[-7pt]
LO & $\vec{\bf A}_{{\rm 1N:  \, static}}, \;$
                Eq.~(4.2) & --- & --- \\ [5pt] \hline
&&& \\[-9pt]
NLO & $\vec{\bf A}_{{\rm 1N:  \, static}}, \;$
                Eq.~(4.7)  & \hspace*{1cm} --- &\hspace*{1cm}  --- \\
[5pt] \hline
&&& \\[-9pt]
N$^2$LO& --- & $\vec{\bf A}_{{\rm 2N:  \, 1\pi}}, \;$ Eq.~(5.7)  & --- \\ [2pt]
& & 
$+ \; \vec{\bf A}_{{\rm 2N:  \, cont}}, \;$
    Eq.~(5.8)  &\\ [5pt] \hline
&&& \\[-9pt]
N$^3$LO & $\vec{\bf A}_{{\rm 1N:  \, static}}, \;$
                Eq.~(4.46) & $\vec{\bf A}_{{\rm
                                                 2N:  \, 1\pi}}, \;$
                                                 Eq.~(5.13) 
                                            &  
$\vec{\bf A}_{{\rm 3N:
                                              \, \pi}}, \;$
                                              Eq.~(6.2) 
  \\ [2pt]
& 
$+ \; \vec{\bf A}_{{\rm 1N:\,}1/m, {\rm
    UT^\prime}}, \;$  Eq.~(4.13) &
$+ \;  \vec {\bf A}_{{\rm 2N:\,}1\pi, {\rm
                                                    UT^\prime} } ,
                                                    \;$
                                                    Eq.~(5.23) 
                                            &
$+ \;  \vec {\bf A}_{{\rm
    3N: \, cont}}, \;$ Eq.~(6.6)  \\ [2pt]
& 
$+ \; \vec{\bf A}_{{\rm 1N: \, 1/m^2}} , \;$
  Eq.~(4.18) &
$+ \;  \vec{\bf A}_{{\rm 2N:} \, 1\pi , \, 1/m },
                                                    \;$
                                Eq.~(5.19) & \\ [2pt]
&&  
$+ \;  \vec{\bf A}_{{\rm 2N:} \, 2\pi },
                                                    \;$
                                Eq.~(5.29) & \\ [2pt]
&&  
$+ \;    \vec {\bf A}_{{\rm 2N:\,cont,\,UT^\prime}} ,
                                                    \;$
                                Eq.~(5.43) & \\
  [2pt]
&&  
$+ \;    \vec {\bf A}_{{\rm 2N:\,cont,\, 1/m}} ,
                                                    \;$
                                Eq.~(5.41) &  \\ &&&
  \\[-11pt]
\hline \hline
\end{tabularx}
\end{table}

\begin{table}
\caption{Chiral expansion of the nuclear axial charge operator up to N$^3$LO.  LO, NLO, N$^2$LO and N$^3$LO refer to chiral orders $Q^{-3}$,
  $Q^{-1}$, $Q^{0}$ and $Q$, respectively. All equation references are understood to be from~\cite{Krebs:2016rqz}. 
\label{tab_sum_ax_charge}}
\smallskip
\begin{tabularx}{\textwidth}{lrrr}
\noalign{\smallskip}
\hline \hline
 order &  single-nucleon  &\hspace*{1.8cm}  two-nucleon  &
                                                            \hspace*{1.8cm}
                                            three-nucleon  
\smallskip
 \\
\hline \hline
&&& \\
LO & ---  & --- & --- \\ [5pt] \hline
&&& \\
NLO & ${\bf A}_{{\rm 1N: \,UT^\prime}}^{0} , \; $
                Eq.~(4.4)  & ${\bf A}^{0}_{{\rm 2N:}\,
  1\pi}, \;$ Eq.~(5.3)  & --- \\ [2pt]
& \hskip -0.35 true cm $+ \; {\bf A}_{{\rm 1N: \,}1/m}^{0} , \;$
  Eq.~(4.10) &  & \\ [5pt]
\hline
&&& \\[-9pt]
N$^2$LO &---  & ---  &---   \\ [5pt] \hline
&&& \\[-9pt]
N$^3$LO & ${\bf A}_{{\rm 1N:\,static, \,UT^\prime}}^{0} , \;$
                Eq.~(4.14) & ${\bf
                                                             A}^{0
                                                             }_{{\rm
                                                             2N:}\,
                                                             1\pi},
                                                             \;$
                                                             Eq.~(5.14)
                                          & ---\\ [2pt]
& 
\hskip -0.35 true cm 
$+ \;  {\bf A}_{{\rm 1N:\,} 1/m}^{0 }    , \;$ Eq.~(4.22) &
\hskip -0.35 true cm 
$+ \;  {\bf A}^{0}_{{\rm 2N:}\, 2\pi}   , \;$
                                                                                                               Eq.~(5.30) &\\ [2pt]
&&
\hskip -0.35 true cm 
$+ \;  {\bf A}^{0}_{\rm 2N:\, cont}  , \;$
                                                                                                               Eq.~(5.34)
                          & \\ &&& \\[-11pt]
\hline \hline
\end{tabularx}
\end{table}

\section{Electroweak Currents within UT vs TOPT}
As already mentioned in the introduction, in parallel to our activities
within UT techniques electroweak currents have been calculated within
TOPT technique  by Pisa-JLab
group~\cite{Baroni:2015uza,Piarulli:2012bn,Pastore:2008ui,Pastore:2009is}. The current
operators in both calculations should agree with each other modulo
unitary transformation. For vector currents it has been shown that
there exists a unitary transformation which transforms UT currents
into TOPT currents~\cite{Pastore:2009is,Pastore:2011ip}. The situation is more complicated for the axial
vector currents. In this case the UT and TOPT results disagree even at
the point of vanishing momentum transfer. An extensive discussion on
 this issue can be found in~\cite{Baroni:2018fdn}. It remains to be seen in the
 future if the currents are unitary equivalent. If this transformation
 exists it should depend explicitly on the axial vector sources. The
 reason is that the TOPT current satisfies (at least in the chiral
 limit) an ordinary continuity equation~\cite{Baroni:2015uza} 
\begin{eqnarray}
\big[W,\tilde{\bf V}_0(\vec{k},0) \big] &=&\vec{k}\cdot\vec{\tilde{\bf
  V}}(\vec{k},0), \nonumber\\ 
\big[W,\tilde{\bf A}_0(\vec{k},0) \big] &=&\vec{k}\cdot\vec{\tilde{\bf
  A}}(\vec{k},0)-m_q\,i\, \tilde{\bf P}(\vec{k},0), 
\label{ordinarycontinuityeqmomspacevector}
\end{eqnarray} 
which means that TOPT currents
 do not depend on energy transfer. Since our currents do depend on the
 energy transfer they satisfy continuity equations in the form of
 Eq.~(\ref{ordinarycontinuityeqmomspacevector}) and can only be
 transformed to TOPT currents (if possible) with source dependent
 unitary transformations.
\section{Towards Consistent Regularization of the Currents}
Sofar all reported calculations of current operators have been
performed by using dimensional regularization. Naively one could take
these operators and start to look at their expectation values in order
to study observables. This is indeed what has been done by various
calculation with TOPT currents, see e.g. \cite{Riska:2016cud} for a
review. All these calculations should be considered as a hybrid
approach where no claim on consistency between nuclear forces and
currents is made. Even if both nuclear forces and currents are
calculated from the same framework of chiral EFT the use of different
regularizations (cut off vs dimensional regularization) leads to a chiral symmetry violation in the very first
iteration of the current with nuclear forces. Here is the
explanation: 

In order to
solve the Schr\"odinger equation nuclear forces have to be
regularized. The usual way is to use the cutoff regularization. Let
us for example choose a semi-local regulator discussed
in~\cite{Reinert:2017usi}. The regularized form of the long-range part of the leading order
nuclear force, which is one pion exchange diagram, is given by 
\begin{eqnarray}
V_{1\pi,\Lambda}&=&-\frac{g_A^2}{4 F_\pi^2}{\bf \tau}_1\cdot{\bf \tau_2}\frac{\vec{\sigma}_1\cdot\vec{q}\, \vec{\sigma}_2\cdot\vec{q}}{q^2+M_\pi^2}e^{-\frac{q^2+M_\pi^2}{\Lambda^2}},
\end{eqnarray}
where $\vec{q}$ denotes momentum transfer between two nucleons. The
nice property of this regulator is that it does not affect long range
part of the nuclear force at any power of $1/\Lambda$. On  the
other hand a pion-pole contribution proportional to $g_A$ of the relativistic correction of 
the axial vector two-nucleon current is given by
\begin{eqnarray}
\label{Current1piRel}
\vec {\bf A}_{{\rm 2N:} \, 1\pi , \, 1/m}^{ (Q,g_A)}&=&\frac{g_A}{8
                                                        F_\pi^2
                                                        m}i \,
                                                        {\bf \tau}_1 \times
{\bf \tau}_2 
\frac{\vec{q}_1\cdot\vec{\sigma}_1}{q_1^2+M_\pi^2}
 \frac{\vec{k}}{k^2+M_\pi^2}\Big[
i\,
\vec{k}\cdot\vec{q}_1\times\vec{\sigma}_2-\vec{k}_1\cdot\vec{q}_1+\vec{k}_2\cdot(\vec{q}_1
                                                        + \vec{k})
\Big]\nonumber\\
&+&
1 \; \leftrightarrow \; 2\,,
\end{eqnarray}
where $\vec{k}$ is the momentum transfer of the axial vector current,
and other momenta are defined by
\begin{eqnarray}
\vec{q}_i&=&\vec{p}_i^\prime-\vec{p}_i, \quad
             \vec{k}_i\,=\,\frac{\vec{p}_i^\prime +\vec{p}_i}{2},
             \quad i\,=\,1,2,
\end{eqnarray}
and momenta $\vec{p}_i^\prime$ and $\vec{p}_i$ correspond to the final
and initial momenta of the $i$-th nucleon, respectively. Note that this
is not the only contribution to the relativistic corrections of the
current, but only that which is proportional to $g_A$. Complete
expression (including terms proportional to $g_A^3$) for the
relativistic corrections can be found in~\cite{Krebs:2016rqz}. After
we regularized the nuclear force and the axial vector current we can
perform the first iteration and take $\Lambda\to\infty$ limit:
\begin{align}
\vec {\bf A}_{{\rm 2N:} \, 1\pi , \, 1/m}^{
  (Q,g_A)}\frac{1}{E-H_0+i\epsilon}V_{1\pi,\Lambda} +
  V_{1\pi,\Lambda}\frac{1}{E-H_0+i\epsilon}
\vec {\bf A}_{{\rm 2N:} \, 1\pi , \, 1/m}^{
  (Q,g_A)}\,=\,\nonumber\\
\Lambda\frac{g_A^3}{32\sqrt{2}\pi^{3/2}F_\pi^4}({\bf
              \tau}_1-{\bf
   \tau}_2)\frac{\vec{k}}{k^2+M_\pi^2}\vec{q}_1\cdot\vec{\sigma}_1+1
   \; \leftrightarrow \; 2\, + {\cal O}(\Lambda^0).\label{firstiteration:axialvectorcurrent}
\end{align}
Since the one loop amplitude should be renormalizable there should
exist a counter term which absorbs the linear singularity in
$\Lambda$. From Eq.~(\ref{firstiteration:axialvectorcurrent}) we see
that this should be a contact two-nucleon interaction with one pion
coupling to it. However, there is no counter term like this in 
chiral EFT. Such counter term requires derivative-less
coupling of the pion which is forbidden by the chiral symmetry: There
exists only a
counter term proportional to $\vec{k}\cdot\vec{\sigma}_1$, but there
is none which is proportional to $\vec{q}_1\cdot\vec{\sigma}_1$. Here
$\vec{k}$ is the momentum of the pion coupling to the two-nucleon
interaction.\footnote{At higher orders one can construct derivative-less
  pion-four-nucleon interactions by multiplying low energy constants
  with $M_\pi^{2}$. They are coming from explicit chiral symmetry
  breaking by finite quark mass. However, at the order $Q$ we can not
  construct a counter term like this.} If there is no counter term
which absorbs the linear cutoff singularity there should be some
cancelation in the amplitude with other terms. Indeed the same
singularity but with opposite sign we would get for the static limit
of the axial vector current of the 
order $Q$ if we would calculate the current by using cutoff
regularization. Axial vector current at the order $Q$, however, is
calculated by using dimensional regularization and is finite. It also
remains finite if we just multiply the current with any cutoff
regulator we want. So at the level of the amplitude the mismatch
between the cutoff and the dimensional regularization used in the
construction of  operators leads to a
violation of chiral symmetry at one-loop level which, however, is the order of
accuracy of our calculations.
So we see that it is dangerous to
multiply the current operators calculated within dimensional
regularization by some cutoff regulator and calculate expectation
values of this. With the similar arguments one
  can show that dimensionally regularized three-nucleon  forces at the
  level of N$^3$LO, which were published in~\cite{Bernard:2007sp,
    Bernard:2011zr}, can not be used in combination with the cutoff regularized
  two-nucleon forces at the same order. The mismatch between dimensional and cutoff
  regularization will lead also in this case to a violation of the
  chiral symmetry at the  one
  loop level.

In order not to violate the chiral symmetry we need to
calculate both nuclear forces and currents with the same regulator. On
top of it the regulator which we choose should be symmetry preserving. One
possibility to construct a regulator, which manifestly respects the chiral
symmetry,  was  proposed more than four decades ago by Slavnov~\cite{Slavnov:1971aw},
where he introduced a so called higher derivative regularization in a
study of non-linear sigma model. Recently, first applications of this technique
to the chiral EFT have been discussed in the
literature~\cite{Djukanovic:2004px, Long:2016vnq}. Construction of
consistent nuclear forces and currents within a 
similar approach is work in progress.


\begin{thebibliography}{99}
\bibitem{Weinberg:1991um} 
  S.~Weinberg,
  Nucl.\ Phys.\ B {\bf 363}, 3 (1991).
  doi:10.1016/0550-3213(91)90231-L

\bibitem{Epelbaum:2008ga} 
  E.~Epelbaum, H.~W.~Hammer and U.~G.~Mei{\ss}ner,
  Rev.\ Mod.\ Phys.\  {\bf 81}, 1773 (2009)
  doi:10.1103/RevModPhys.81.1773
  [arXiv:0811.1338 [nucl-th]].

\bibitem{Reinert:2017usi} 
  P.~Reinert, H.~Krebs and E.~Epelbaum,
  Eur.\ Phys.\ J.\ A {\bf 54}, no. 5, 86 (2018)
  doi:10.1140/epja/i2018-12516-4
  [arXiv:1711.08821 [nucl-th]].

\bibitem{Park:1995pn} 
  T.~S.~Park, D.~P.~Min and M.~Rho,
  Nucl.\ Phys.\ A {\bf 596}, 515 (1996)
  doi:10.1016/0375-9474(95)00406-8
  [nucl-th/9505017].

\bibitem{Park:1993jf} 
  T.~S.~Park, D.~P.~Min and M.~Rho,
  Phys.\ Rept.\  {\bf 233}, 341 (1993)
  doi:10.1016/0370-1573(93)90099-Y
  [hep-ph/9301295].

\bibitem{Krebs:2016rqz}
  H.~Krebs, E.~Epelbaum and U.-G.~Mei{\ss}ner,
  ``Nuclear axial current operators to fourth order in chiral effective field theory,''
  Annals Phys.\  {\bf 378} (2017) 317

\bibitem{Krebs:2019aka}
  H.~Krebs, E.~Epelbaum and U.-G.~Mei{\ss}ner,
  ``Nuclear electromagnetic currents to fourth order in chiral effective field theory,''
  arXiv:1902.06839 [nucl-th].

\bibitem{Baroni:2015uza}
  A.~Baroni, L.~Girlanda, S.~Pastore, R.~Schiavilla and M.~Viviani,
  ``Nuclear Axial Currents in Chiral Effective Field Theory,''
  Phys.\ Rev.\ C {\bf 93} (2016) no.1,  015501
   Erratum: [Phys.\ Rev.\ C {\bf 93} (2016) no.4,  049902]
   Erratum: [Phys.\ Rev.\ C {\bf 95} (2017) no.5,  059901]

\bibitem{Piarulli:2012bn}
  M.~Piarulli, L.~Girlanda, L.~E.~Marcucci, S.~Pastore, R.~Schiavilla and M.~Viviani,
  ``Electromagnetic structure of A = 2 and 3 nuclei in chiral effective field theory,''
  Phys.\ Rev.\ C {\bf 87} (2013) no.1,  014006

\bibitem{Pastore:2008ui} 
  S.~Pastore, R.~Schiavilla and J.~L.~Goity,
  Phys.\ Rev.\ C {\bf 78}, 064002 (2008)
  doi:10.1103/PhysRevC.78.064002
  [arXiv:0810.1941 [nucl-th]].

\bibitem{Pastore:2009is} 
  S.~Pastore, L.~Girlanda, R.~Schiavilla, M.~Viviani and R.~B.~Wiringa,
  Phys.\ Rev.\ C {\bf 80}, 034004 (2009)
  doi:10.1103/PhysRevC.80.034004
  [arXiv:0906.1800 [nucl-th]].

\bibitem{Pastore:2011ip} 
  S.~Pastore, L.~Girlanda, R.~Schiavilla and M.~Viviani,
  Phys.\ Rev.\ C {\bf 84}, 024001 (2011)
  doi:10.1103/PhysRevC.84.024001
  [arXiv:1106.4539 [nucl-th]].

\bibitem{Slavnov:1971aw} 
  A.~A.~Slavnov,
  Nucl.\ Phys.\ B {\bf 31}, 301 (1971).
  doi:10.1016/0550-3213(71)90234-3

\bibitem{Kolling:2011mt} 
  S.~K\"olling, E.~Epelbaum, H.~Krebs and U.-G.~Mei{\ss}ner,
  Phys.\ Rev.\ C {\bf 84}, 054008 (2011)
  [arXiv:1107.0602 [nucl-th]].

\bibitem{Kolling:2009iq} 
  S.~K\"olling, E.~Epelbaum, H.~Krebs and U.-G.~Mei{\ss}ner,
  Phys.\ Rev.\ C {\bf 80}, 045502 (2009)
  [arXiv:0907.3437 [nucl-th]].

\bibitem{Baroni:2018fdn} 
  A.~Baroni {\it et al.},
  Phys.\ Rev.\ C {\bf 98}, no. 4, 044003 (2018)
  doi:10.1103/PhysRevC.98.044003
  [arXiv:1806.10245 [nucl-th]].

\bibitem{Riska:2016cud} 
  D.~O.~Riska and R.~Schiavilla,
  Int.\ J.\ Mod.\ Phys.\ E {\bf 26}, no. 01n02, 1740022 (2017)
  doi:10.1142/S0218301317400225
  [arXiv:1603.01253 [nucl-th]].

\bibitem{Djukanovic:2004px} 
  D.~Djukanovic, M.~R.~Schindler, J.~Gegelia and S.~Scherer,
  Phys.\ Rev.\ D {\bf 72}, 045002 (2005)
  doi:10.1103/PhysRevD.72.045002
  [hep-ph/0407170].

\bibitem{Long:2016vnq} 
  B.~Long and Y.~Mei,
  Phys.\ Rev.\ C {\bf 93}, no. 4, 044003 (2016)
  doi:10.1103/PhysRevC.93.044003
  [arXiv:1605.02153 [nucl-th]].

\bibitem{Bernard:2007sp} 
  V.~Bernard, E.~Epelbaum, H.~Krebs and U.~G.~Mei{\ss}ner,
  Phys.\ Rev.\ C {\bf 77}, 064004 (2008)
  doi:10.1103/PhysRevC.77.064004
  [arXiv:0712.1967 [nucl-th]].

\bibitem{Bernard:2011zr} 
  V.~Bernard, E.~Epelbaum, H.~Krebs and U.-G.~Mei{\ss}ner,
  Phys.\ Rev.\ C {\bf 84}, 054001 (2011)
  doi:10.1103/PhysRevC.84.054001
  [arXiv:1108.3816 [nucl-th]].
\end{thebibliography}
\end{document}